\documentclass[12pt]{iopart}
\usepackage{txfonts}
\usepackage{graphicx}
\usepackage{iopams,setstack}
\usepackage{cite}
\begin{document}

\title[]{Depth penetration and scope extension of failures in the cascading of multilayer networks}

\author{Wen-Jun Jiang$^{1}$, Run-Ran Liu$^{1}$ and Chun-Xiao Jia$^{1}$}
\address{$^1$ Research Center for Complexity Sciences, Hangzhou Normal University, Hangzhou, Zhejiang 311121, China}

\begin{abstract}
Real-world complex systems always interact with each other, which causes these systems to collapse in an avalanche or cascading manner in the case of random failures or malicious attacks. The robustness of multilayer networks has attracted great interest, where the modeling and theoretical studies of which always rely on the concept of multilayer networks and  percolation methods. A straightforward and tacit assumption is that the interdependence across network layers is strong, which means that a node will fail entirely with the removal of all links if one of its interdependent neighbours fails. However, this oversimplification cannot describe the general form of interactions across the network layers in a real-world multilayer system. In this paper, we reveal the nature of the avalanche disintegration of general multilayer networks with arbitrary interdependency strength across network layers. Specifically, we identify that the avalanche process of the whole system can essentially be decomposed into two microscopic cascading dynamics in terms of the propagation direction of the failures: depth penetration and scope extension. In the process of depth penetration, the failures propagate from layer to layer, where the greater the number of failed nodes is, the greater the destructive power that will emerge in an interdependency group. In the process of scope extension, failures propagate with the removal of connections in each network layer. Under the synergy of the two processes, we find that the percolation transition of the system can be discontinuous or continuous with changes in the interdependency strength across network layers, which means that sudden system-wide collapse can be avoided by controlling the interdependency strength across network layers. Our work not only reveals the microscopic mechanism of global collapse in multilayer infrastructure systems but also provides stimulating ideas on intervention programs and approaches for cascade failures.
\end{abstract}

\date{\today}

\maketitle

\section{Introduction} \label{sec:intro}
Many real-world complex systems, both natural~\cite{Klosik:2017} and man made ~\cite{Ouyang:2014,Rinaldi:2001,Radicchi:2015}, can be described as multilayer or interdependent networks given the existence of different levels of interdependence across network layers. Recent theoretical studies on networks with two or more layers show that when the nodes in each network are interdependent on the nodes in other networks, even small initial failures can propagate back and forth and lead to the abrupt collapse of the whole system~\cite{BPPSH:2010,KABGP:2014,Gao:2011,Gao:2012,Baxter248701:2012}. In this sense, multilayer networks are more fragile than single-layer networks in resisting the propagation of initial failures~\cite{BPPSH:2010}. In recent years, we have witnessed considerable progress in the study of multilayer networks with the aid of percolation theory~\cite{Kesten:1982,Stauffer:1992,Son:2012}. It has been found that multilayer networks are not as fragile as in theoretical studies under certain specific conditions such as those given link overlap~\cite{Cellai:2013,Hu:2013}, geometric correlations~\cite{Kleineberg:2016,Kleineberg:2017}, correlated community structures~\cite{Faqeeh:2018}, inter-layer degree correlations~\cite{Min:2014,Parshani:2011}, intra-layer degree correlations~\cite{Valdez:2013}, and autonomous nodes~\cite{PBH:2010,Schneider:2013,Valdez:2014} being able to facilitate the viability of nodes and alleviate the suddenness of the collapse in an interdependent system. In addition, some real properties facing real interdependent systems, such as spatial constraints~\cite{Berezin:2015,Bashan:2013,Danziger:2014,Shekhtman:2014}, clustering~\cite{Shao:2014,Huang:2013}, and degree distribution~\cite{Emmerich:2014,Yuan:2015}, also enhance the robustness and mitigate cascading failures of interdependent networks.

A key question in the modeling of multilayer networks is how to describe the interdependencies across network layers. A straightforward method employed in most previous models of cascading failures in multilayer networks is assuming that the layer interdependence is ``strong'', where a failure node can cause all of its interdependent neighbors to fail completely~\cite{PBH:2010,BPPSH:2010,Hu:2017PNAS,Shaw:2010,Zhao:2014,NA:2015,WangZexun,Liu:2019}. This assumption has already been extended extensively to the study of cascading dynamics in networks under different conditions such as interdependency groups in single-layer networks~\cite{Bashan2011,Bashan2011b,Liming:2013} and k-core percolation~\cite{Azimi:2014}, weak percolation~\cite{Baxter:2014} and redundant percolation~\cite{Radicchi:2017} in multilayer networks. Nevertheless, this assumption is somewhat simplistic and cannot cover the case where nodes are weakly interdependent. For instance, in a civil transportation system, the flow of passengers from city to city depends on a number of transportation modes  such as coaches, trains, airplanes, and ferries. When any mode becomes unavailable, the total failure of the other three modes seems impossible, e.g., when a local train station is shut down, passenger flow into the city may be decreased: some passengers destined for this city may cancel their trips, and the transferring passengers would switch to other cities to reach their destinations. The reduction of passenger flow can cause some routes in other modes to not operate properly, and carriers experience financial or other losses; for instance, airlines may cancel flights if passenger numbers are below expectations. Specifically, the interdependence across network layers can be ``weak'', and the failure of a node cannot destroy its all of dependency neighbours with probability $1$~\cite{LESL:2018,Liu:2016B}. Under these circumstances, the failure of a train station can cause one or more of its interdependent nodes in other network layers to suffer damage or even failure, e.g., the failure of a local coach station, which can further lead to failures in more modes and deteriorate the connectivity of the city. By this token, there may exist a cascading process underlying a group of interdependent nodes across network layers, which means that the microscopic mechanism of global collapse in multilayer networks could include not only the propagation of failures from node to node inside a certain network layer but also the cascading process of failures across network layers. In this paper, we regard the propagation of failures inside a network layer as ``inner-layer cascading'' and the propagation of failures in a dependency group across network layers as ``cross-layer cascading''.

Previous networks that have considered the ``strong'' interdependence ignore the microscopic process of ``cross-layer cascading'', as the failure of one node will destroy all its interdependent neighbours. In this paper, we aim to model the cascading dynamics in multilayer networks within a more general situation by using the assumption of ``weak'' interdependence~\cite{LESL:2018}, where the strength of interdependence can be tuned by introducing a tolerance parameter $\alpha$. Using a comprehensive theoretical study and numerical simulations, we find that the cascading dynamic in multilayer networks is essentially the synergistic result of ``cross-layer cascading'' and ``inner-layer cascading''. In particular, we find that the system can undergo different types of percolation with changing tolerance parameters $\alpha$. Specifically, the system percolates as an abrupt (first-order) percolation transition for small values of $\alpha$. With increasing $\alpha$ exceeding a critical value $\alpha_{c}$, the system percolates in a continuous (second-order) manner. However, for scale-free networks, the phenomenon of double phase transitions occurs for some moderate parameter values of $\alpha$, where the networks in the system first percolate in a continuous (second-order) manner and then experience a first-order phase transition in an abrupt manner at another phase transition point.

\section{Model} \label{sec:model}
We consider a multilayer network consisting of $M$ layers of networks, where each network layer has  $N$  nodes. We label the network layers with Latin letters $A$, $B$, $C$, $\cdots$, and the nodes in each network layer are labeled with Arabic numbers $1,2, \cdots, N$. Therefore, each node in a certain network layer can be identified as a pair of coordinates ($x$, $X$), with $x$ denoting the node label and $X$ denoting the layer label. The nodes across network layers with the same Arabic number are regarded as replica nodes, and they are interdependent on each other. The nodes in the same network layer $X$ are linked by a set of connectivity links, and the connectivity degree of nodes follows a degree distribution $p_{k}^{X}$.

The cascading in the multilayer networks is triggered by randomly removing a fraction $1-p$ of nodes and their replicas. In each network layer, the links connected to the removed nodes are removed simultaneously, which causes the network layer to break up into a set of connected components~\cite{Molloy:1995}. The nodes in the giant component are regarded as functional, and the other nodes are treated as failed. Due to the interdependency among the replica nodes across network layers, a failed node will further cause a certain level of damage to it replicas, where the damage degree is controlled by the tolerance parameter $\alpha$. Specifically, when one node in a network layer fails, each connectivity link of its viable replicas in other network layers will be disabled with a probability $1-\alpha$. Along with the removal of the connectivity links, the remaining viable replicas can fail due to isolation from the giant components, and the network layers will be further fragmented and thus lead to more failures simultaneously. Specifically, the failures can propagate from layer to layer through the interdependencies among replica nodes, and the failures can also propagate from node to node in a certain network layer in a multilayer system simultaneously. After a number of iterations of link removal caused by node failures and node isolations resulting from network fragmentation, the system can reach a steady state. In this paper, we use the relative size $S^X$ of the giant component in each layer of network $X$ to measure the robustness of the network.

\begin{figure}[htbp]
\centering
\includegraphics[width=0.9\linewidth]{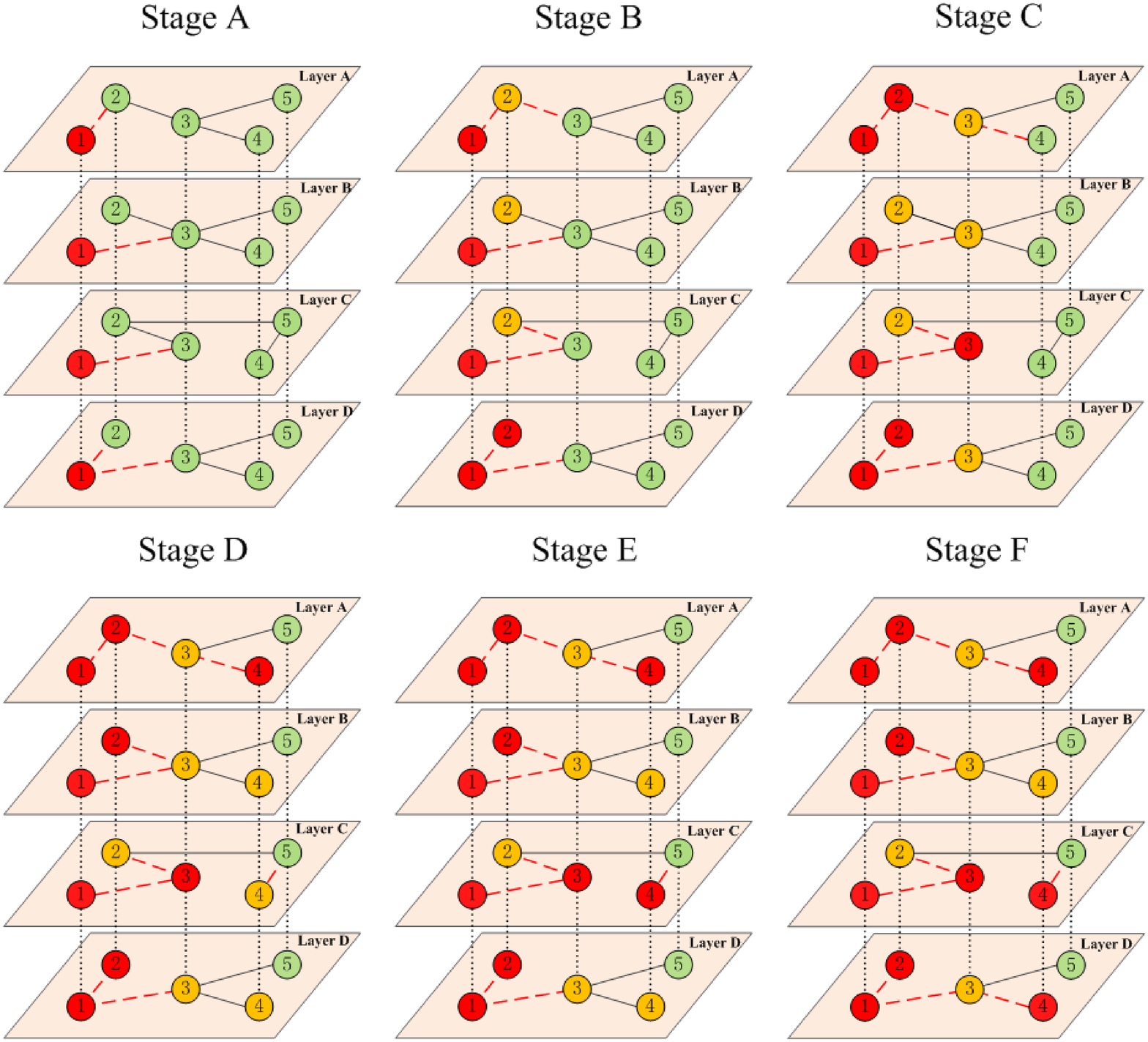}
\caption{Schematic diagram of the cascading process in a four-layer network. The nodes in different layers with the same Arabic number are replica nodes, and they are connected by dotted lines. A dashed line denotes the connectivity link in the network layer. The functional nodes are marked in green, the failed nodes are marked in red, and the yellow nodes are still viable after being damaged. At stage A, the replicas with the Arabic number $1$ are removed from all layers. At stage B, the node ($2$, $D$) becomes isolated from the giant component in layer $D$ and fails, which leads to the link removal of its replicas in layers $A$, $B$ and $C$. At stage C, the node ($3$, $C$) becomes isolated from the giant component in layer $C$ and fails, which leads to the link removal of its replicas in layers $A$, $B$ and $D$. Simultaneously, node ($2$, $A$) becomes isolated and fails due to the removal of link $23$ in layer $A$, which further leads to the link removal of its replicas in layers $B$ and $C$. At stage D, the node ($4$, $A$) becomes isolated from the giant component in layer $A$ and fails, which leads to the link removal of its replicas in layers $B$, $C$ and $D$. Simultaneously, node ($2$, $A$) becomes isolated and fails due to the removal of link $23$ in layer $B$. At stage E, the node ($4$, $C$) becomes isolated from the giant component and fails, which leads to the link removal of its replicas in layers $B$ and $D$. At stage F, the node ($4$, $D$) becomes isolated and fails, and the system reaches the final steady state.}
\label{fig:cascade}
\end{figure}

\section{Theory} \label{sec:theory}
We use the method of probability generation functions~\cite{Son2011,Feng:2015} to obtain the theoretical solution of the model, and the generating function $G^{X}_{0}(x)=\sum_kp^{X}_{k}x^k$ is employed to generate the degree distribution $p^{X}_{k}$ of layer $X$. Similarly, the generating function $G^{X}_{1}(x)=\sum_{k}p^{X}_{k}kx^{k-1}/\langle k\rangle^{X}$ is used to generate the excess degree distribution of a node reached by following a random link, where $\langle k\rangle \equiv \sum_{k}p^{X}_{k}k$ represents the average degree of the network layer $X$. In particular, we define $R^{X}$ as the probability that a randomly chosen link in network $X$ belongs to its giant component in the steady state of the system. For simplicity, we consider the case where the $M$ network layers within the multilayer system have an identical degree distribution: $p_{k}^{X}=p_{k}$. We thus have $G^{X}_{0}(x)\equiv G_{0}(x)$, $G^{X}_{1}(x)\equiv G_{1}(x)$, $R^{X}\equiv R$, and $\langle k\rangle^X\equiv \langle k\rangle$ to simplify the notations. Assuming that each network of $M$ network layers is tree like, we aim to obtain the equation governing the probability $S^{X}$ that a random node is in the giant component of layer $X$. Because each layer has the same degree distribution, we  have $S^{A}=S^{B}=S^{C} \cdots \equiv S$. Following a randomly chosen link in the layer $X$, we arrive at a node $(x,X)$ of degree $k$ with $t$ failed replicas. Therefore, each link of  node $(x,X)$ is preserved with a probability $\alpha^{t}$. Considering that the degree $k$ follows the probability distribution $kp_k/z$, the probability that the random link can lead to the giant component follows $\alpha^{t}[1-kp_k/z(1-\alpha^{t}R)^{k-1}]$, which can be simplified as $\alpha^{t}[1-G_{1}(1-\alpha^{t}R)]$ in terms of the generating function $G_{1}(x)$. If the number $t$ of failed replicas for a given node follows a probability distribution $f(t)$, we can obtain the self-consistent equation for $R$ by summing over all possible $t$
\begin{equation} \label{eq:R}
R=p\sum^{M-1}_{t=0}\alpha^{t}[1-G_{1}(1-\alpha^{t}R)]f(t) \equiv h(R).
\end{equation}
Similarly, we can obtain the probability $S$ that a random node is in the giant component:
\begin{equation} \label{eq:S}
S=p\sum^{M-1}_{t=0}[1-G_{0}(1-\alpha^{t}R)]f(t).
\end{equation}

The solution process of Eqs.~(\ref{eq:R}) and (\ref{eq:S}) utilizes the probability distribution function $f(t)$, which can be obtained by using the probability $R$. Considering that there are $t$ failed replicas for a random node ($x$,$X$) in the layer $X$ at  steady state, the viable probability of each replica is $1-G_{0}(1-\alpha^{t}R)$, and the remaining $M-t-1$ replicas are all viable with  probability $[1-G_{0}(1-\alpha^{t}R)]^{M-t-1}$. Because the failures can propagate from layer to layer through the interdependencies among replica nodes, we assume that there are $s$ failed replicas caused by the link removal of other nodes in the corresponding network layers, and there are $t-s$ failed nodes induced by the $s$ failed replicas. The probability of  $s$ failed replicas existing caused by isolation is $G^{s}_{0}(1-R)$. After that, the probability of $t-s$ additional replicas failing is $[G_{0}(1-\alpha^{s}R)-G_{0}(1-R)]^{t-s}$. Therefore, $f(t)$ satisfies
\begin{equation}\label{eq:ft}
\eqalign{
f(t)= & \left(
\begin{array}{c}
M-1 \\
t \\
\end{array}
\right)[1-G_{0}(1-\alpha^{t}R)]^{M-t-1} \cr
& \sum^{t}_{s=0}
\left(
\begin{array}{c}
t \\
s \\
\end{array}
\right)G^{s}_{0}(1-R)[G_{0}(1-\alpha^{s}R)-G_{0}(1-R)]^{t-s}.
}
\end{equation}
For a given degree distribution $p_{k}$, we can obtain the final size $S$ of the giant component in a certain layer by solving Eqs.~(\ref{eq:R}) and ~(\ref{eq:S}) simultaneously.

When $\alpha \rightarrow 1$, the interdependence across network layers is weakest, and the system percolates in a second-order manner as in single-layer networks ~\cite{CEAH:2000,CNSW:2000}. When $\alpha \rightarrow 0$, the interdependence across network layers is the strongest, and the system percolates in a first-order manner~\cite{Baxter248701:2012}. Therefore, the manner of percolation transitions can be determined by the value of $\alpha$, and the percolation transition of the system can switch from a second-order to a first-order  percolation when $\alpha$ exceeds a critical value $\alpha_{c}$. For the second-order percolation transition, the probability $R$ tends to zero when $p$ is close to the second-order percolation point $p^{II}_{c}$. We can use the Taylor expansion of Eq.~(\ref{eq:R}) for $R \equiv \epsilon \rightarrow 0$ and $p \rightarrow p^{II}_{c}$:
\begin{equation} \label{eq:R1}
h(\epsilon)= h'(0)\epsilon+\frac{1}{2}h''(0)\epsilon^{2}+O(\epsilon^{3})=\epsilon.
\end{equation}
Ignoring the high-order terms of $\epsilon$, we  have $h'(0)=1$ when $p \rightarrow p^{II}_{c}$. We thus have the condition for the second-order percolation transitions
\begin{equation} \label{eq:CP2}
p^{II}_{c}\alpha^{2M-2}G'_{1}(1)=1,
\end{equation}
and the second-order percolation point
\begin{equation} \label{eq:CPP2}
p^{II}_{c}=\frac{1}{\alpha^{2M-2}G'_{1}(1)}.
\end{equation}
When $\alpha=1$ or $M=1$, the second-order percolation transition point $p^{II}_{c}=\frac{1}{G'_{1}(1)}$, which is coincident with the result for the ordinary percolation in a single-layer network.

At the first-order phase transition point $p^{I}_{c}$, the probability $R$ jumps from $R_{c}$ to zero or a nontrivial value, and the curve of $y=h(R)-R$ is tangent with the straight line $y=0$:
\begin{equation} \label{eq:CP1}
\frac{dh(R)}{dR}|_{R=R_{c},p=p_{c}^{I}}=1.
\end{equation}
In this paper, we resort to the numerical method to solve Eq.~(\ref{eq:CP1}) and Eq.~(\ref{eq:R}) for the first-order percolation transition point $p^{I}_{c}$.

When $\alpha=\alpha_{c}$, the conditions for the first- and second-order percolation transitions are satisfied simultaneously, i.e., $p^{I}_{c}=p^{II}_{c}$ for $R_{c}\rightarrow 0$ at the percolation transition point. At this time, Eq.~(\ref{eq:R1}) reduces to
\begin{equation} \label{eq:R2}
\frac{1}{2}h''(0)\epsilon^{2}+O(\epsilon^{3})=0.
\end{equation}
Therefore, we know that $h''(0)=0$ when $p \rightarrow p_{c}$ and $\alpha=\alpha_{c}$. We can have
\begin{equation} \label{eq:R3}
-\alpha^{3}G''_{1}(1)-2\alpha^{2}(M-1)G'_{1}(1)G'_{0}(1)+2(M-1)G'_{1}(1)G'_{0}(1)=0.
\end{equation}
By the solution of Eq.~(\ref{eq:R3}), we can obtain the switch point $\alpha_{c}$ of first-order and second-order percolation transitions.

\begin{figure}[htbp]
\centering
\includegraphics[width=\linewidth]{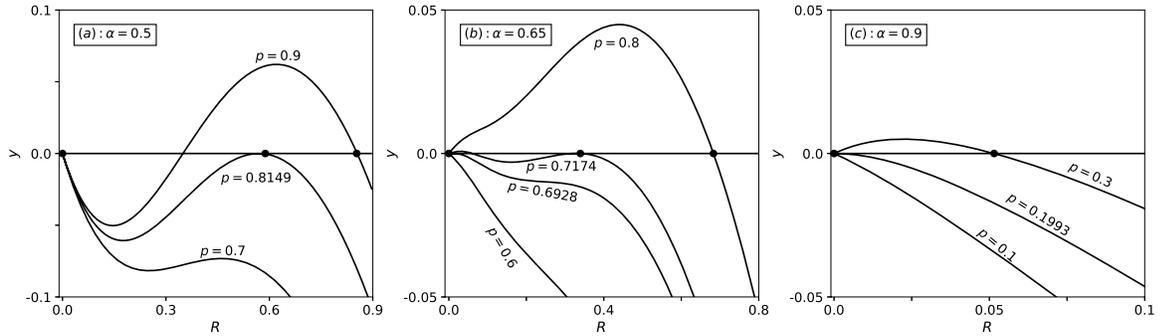}
\caption{ The graphical solutions of Eq.~(\ref{eq:R}) for  different values of $p$ and $\alpha$, as marked by the black dots. (a) the result for $\alpha=0.5$, (b)  the result for $\alpha=0.65$, and (c)  the result for $\alpha=0.9$. For each panel, the degree distribution of networks follows a truncated power-law distribution $p_k\sim k^{-\gamma}(k_{min}\leq k\leq k_{max})$, where $k_{min}=2$, $k_{max}=141$, and $\gamma=2.3$. }
\label{figure2}
\end{figure}

Figure~\ref{figure2} shows the function curves $y=h(R)-R$ for different values of $p$, from which we can validate the existence of first- and second-order percolation transitions. Fig.~\ref{figure2}(a) shows the graphical solutions of Eq.~(\ref{eq:R}) for $\alpha < \alpha_{c}$, from which we can find that the solution of $R$ is given by the tangent point when $p=p^{I}_{c}$, indicating a discontinuous percolation transition. From Fig.~\ref{figure2}(c), we can find that the nontrivial solution of $R$ emerges at the point $p=p^{II}_{c}$, at which the function curve $y=h(R)-R$ is tangent with the $R$ axis at $R=0$, indicating a continuous percolation transition. Interestingly, we also found that the system undergoes double phase transitions for some moderate values of $\alpha$ if the degree distribution $p_{k}$ of the systems is scale free, as shown in Fig.~\ref{figure2}(b). In this case, the system first percolates in a second-order manner and then experiences a first-order phase transition with increased $p$, and the conditions~(\ref{eq:CP2}) and~(\ref{eq:CP1}) should be satisfied at the phase transition points $p^{II}_{c}$ and $p^{I}_{c}$ successively. If the condition~(\ref{eq:CP2}) cannot be satisfied, the double phase transition reduces to a single first-order percolation transition, and if the condition~(\ref{eq:CP1}) cannot be satisfied, the double phase transition reduces to a single second-order percolation transition. Using these conditions, we can locate the boundary between double phase transitions and single first-order phase transitions and the boundary between double phase transitions and single first-order percolation transitions.

\section{Results} \label{sec:results}
\subsection{Synthetic network}
\begin{figure}[htbp]
\centering
\includegraphics[width=\linewidth]{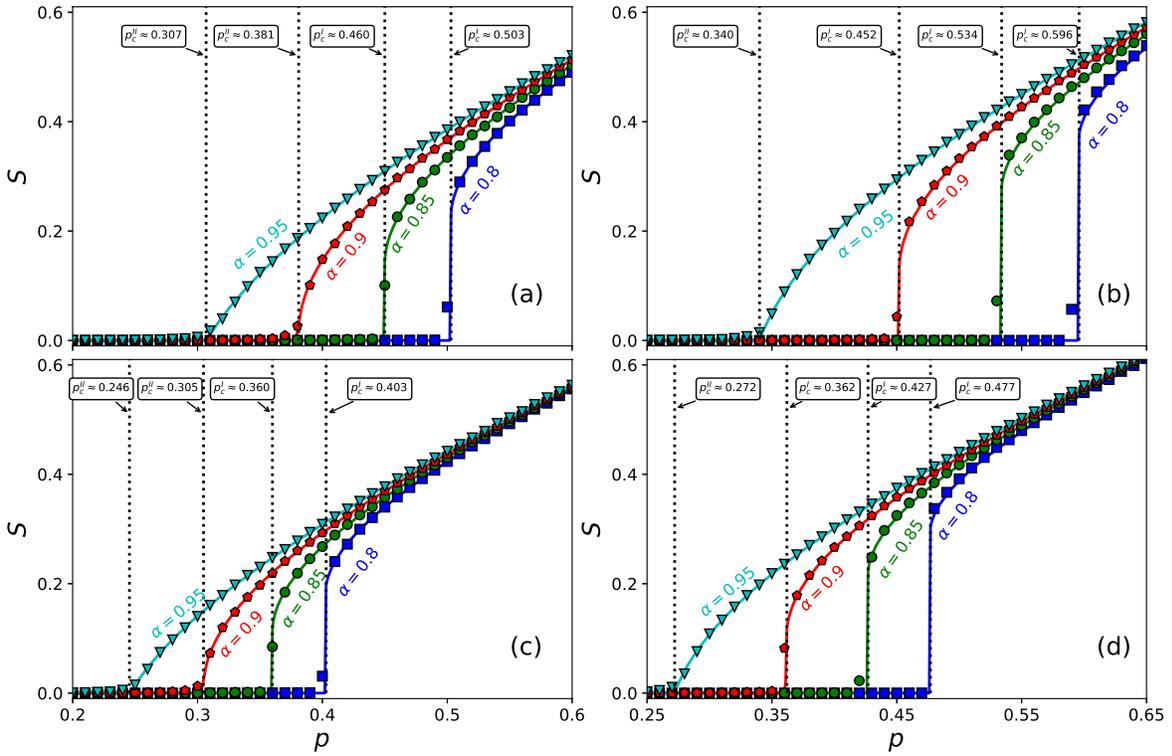}
\caption{ Simulation results for percolation transitions on three-layer and four-layer random networks. (a) and (b) are the results for three-layer and four-layer random networks with $\langle k\rangle=4$, respectively. (c) and (d) are the results for three-layer random networks and four-layer random networks with $\langle k\rangle=5$, respectively. The results were obtained by averaging over $100$ independent realizations, and the network size was $N=10^{5}$. The solid lines behind the symbols denote the theoretical predictions that were obtained by Eqs.~(\ref{eq:R}) and (\ref{eq:S}). The vertical dashed lines denote the first- and second-order percolation transition points predicted by Eqs. (\ref{eq:CP1}) and (\ref{eq:CP2}), respectively. }
\label{fig:ER}
\end{figure}

In this paper, we take two special multilayer networks with  $M=3$ and $M=4$  layers as examples to illustrate the characteristics of percolation dynamics. Figures~\ref{fig:ER} (a) and (b) show the size $S$ of the giant component as functions of $p$ for different values of $\alpha$ in three-layer random networks and four-layer random networks with $\langle k \rangle=4$, respectively. For both three-layer and four-layer random networks, we find that the system can percolate in a discontinuous manner for a small value of $\alpha$ or in a continuous manner for a large value of $\alpha$. In addition, the percolation transition point of three-layer networks is less than that of four-layer networks, which means that three-layer networks are always more robust than four-layer networks. Simultaneously, we find that the critical point $\alpha_{c}$ separating the types of percolation transitions depends on the number of layers in the system. Similar results can also be found for three-layer random networks and four-layer random networks with $\langle k \rangle=5$. In Fig.~\ref{fig:ER}, the theoretical predictions are also provided and agree with the simulation results very well.

\begin{figure}[htbp]
\centering
\includegraphics[width=\linewidth]{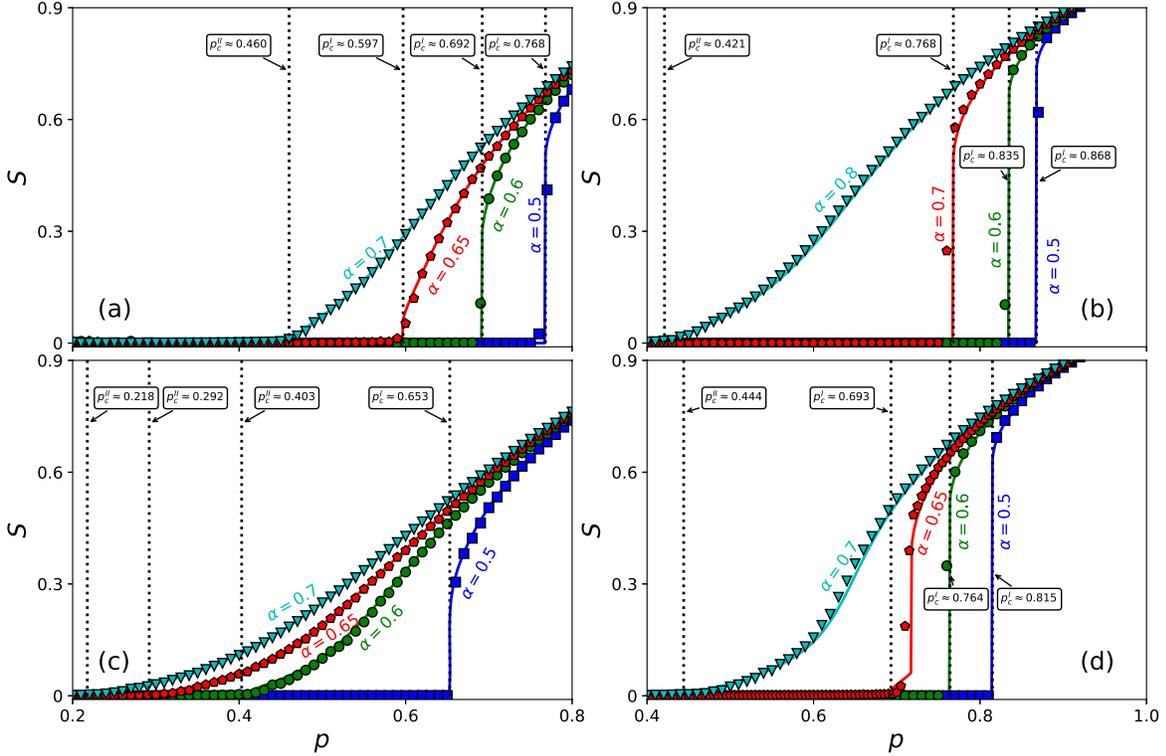}
\caption{Simulation results for percolation transitions on three- and four-layer scale-free networks. (a) and (b) are the results for three- and four-layer scale-free networks with $\langle k\rangle=4$, respectively. (c) and (d) are the results for three- and four-layer scale-free networks with $\langle k\rangle=5$, respectively. When the average degree $\langle k\rangle=4$, the parameter settings for the power-law distribution are $k_{min}=2$, $k_{max}=63$ and $\gamma=2.5$. When the average degree $\langle k\rangle=5$, the parameter settings are $k_{min}=2$, $k_{max}=141$ and $\gamma=2.3$. The simulation results were obtained by averaging over $100$ independent realizations, and the network size is $N=10^{5}$. The solid lines behind the symbols denote the theoretical predictions that were obtained by Eqs.~(\ref{eq:R}) and (\ref{eq:S}). The vertical dashed lines denote the first- and second-order percolation transition points predicted by Eqs. (\ref{eq:CP1}) and (\ref{eq:CP2}), respectively.}
\label{fig:SF}
\end{figure}

The results for multilayer scale-free networks are also provided in Figure~\ref{fig:SF}. For each network layer of the system, the degree of the nodes follows a truncated power-law distribution with an average $\langle k\rangle$, and the degree distribution is $p_k\sim k^{-\gamma}(k_{min}\leq k\leq k_{max})$, where $k_{min}$ and $k_{max}$ are the lower and upper bounds of the degree, respectively, and $\gamma$ is the power law exponent. Similarly, we can also find that the system can percolate in a discontinuous manner for a small value of $\alpha$ or in a continuous manner for a large value of $\alpha$, and the three-layer networks are more robust than four-layer networks for both $\langle k\rangle=4$ and $\langle k\rangle=5$. Interestingly, we can also find that the four-layer networks can undergo a double phase transition for $\alpha=0.65$. Specifically, the system first percolates as a continuous phase transition and then undergoes a discontinuous phase transition with increasing $p$. With increasing $\alpha$, the discontinuous phase transition disappears, and the system reduces to a single continuous phase transition. With decreasing $\alpha$, the continuous phase transition disappears, and the system reduces to a single discontinuous phase transition.

\begin{figure}[htbp]
\centering
\includegraphics[width=\linewidth]{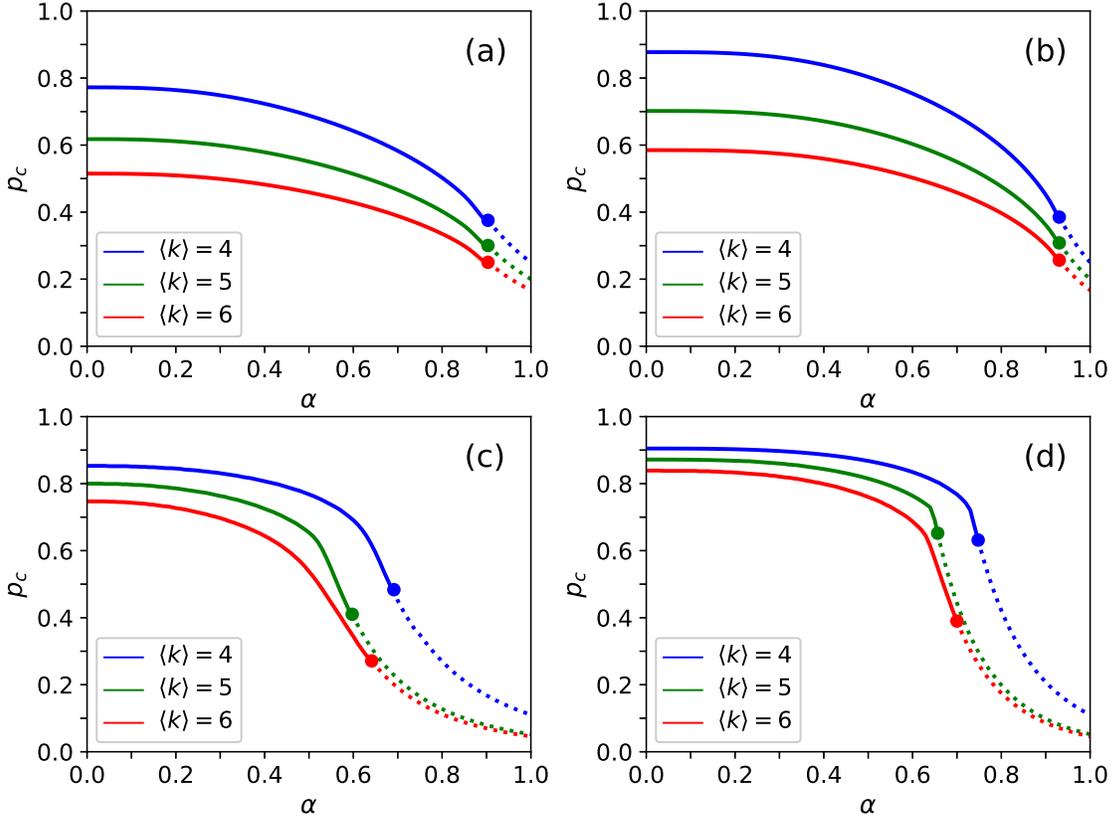}
\caption{(Color online.) (a) The percolation transition point $p_{c}$ ($(p_{c}^{I})$ or $(p_{c}^{II})$) versus $\alpha$ for three-layer random networks, where the average degree $\langle k\rangle$ is $4$, $5$ and $6$. (b) Corresponding results for four-layer random networks. (c)  the corresponding results for three-layer scale-free networks of average degree $\langle k\rangle$ $4$, $5$ and $6$ (corresponding to a power-law exponent of degree distribution $-2.6$, $-2.3$, and $-2.1$, respectively) with minimum degree $2$. (d) Corresponding results for four-layer scale-free networks. }
\label{fig:transition}
\end{figure}

Figures~\ref{fig:transition} (a) and (b) show the percolation transition points $p_{c}$ as functions of $\alpha$ for three- and four-layer random networks with different average degrees, respectively. For both three- and four-layer random networks, the manners of percolation transition are classified as discontinuous and continuous by a critical value of $\alpha_{c}$, and the critical value of $\alpha_{c}$ only depends on the number of network layers  $M$  in a system. Figures~\ref{fig:transition} (c) and (d) show the percolation transition points $p_{c}$ as functions of $\alpha$ for three- and four-layer scale-free networks with different average degrees, respectively, from which we can also find that the manners of percolation transition are classified as discontinuous and continuous by a critical value of $\alpha_{c}$; however, the specific value of $\alpha_{c}$ depends on the parameter settings of the degree distributions.

\subsection{Empirical networks}
\begin{figure}[htbp]
\centering
\includegraphics[width=\linewidth]{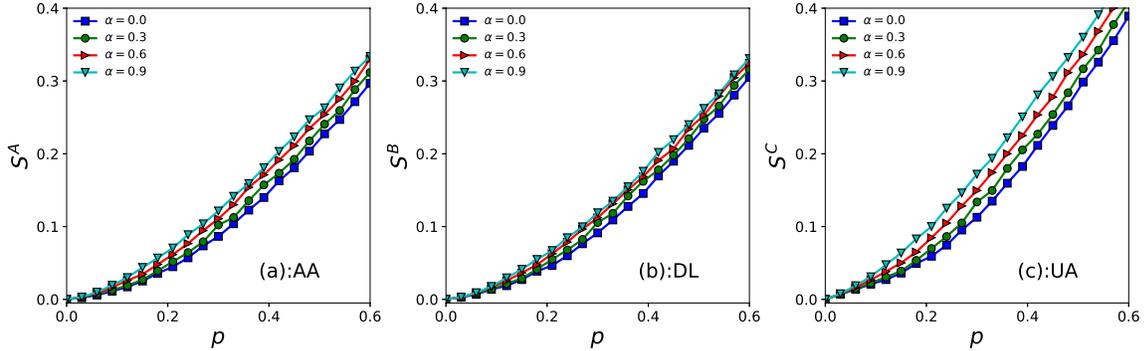}
\caption{ Percolation in multilayer empirical networks. The system consists of the three layers of networks A, B and C, which represent three major carriers: American Airlines (AA), Delta Air Lines (DL), and United Airlines (UA), respectively. (a-c) The sizes of the giant components of the network layers $A$, $B$, and $C$ as  functions of $p$ for different values of $\alpha$, respectively. The data points are the result of averaging over $1000$ statistical realizations. }
\label{fig:robust}
\end{figure}

To address the percolation process in empirical multilayer networks, we consider a three-layer system constituting the three major carriers in the United States: American Airlines (AA), Delta Air Lines (DL), and United Airlines (UA). In each layer of a network, airports are nodes, and connections in the layers are determined by the existence of at least one flight operated by a given carrier between  two airports. We construct the multilayer system using the dataset from OpenFlights (https://openflights.org/data.html). For civil flights of the three major carriers, there are in total $N = 310$ nodes (functioning airports). Some of the nodes do not appear as connected in all the layers, leading to a difference in the relative sizes of the giant components. Figure~\ref{fig:robust}(a-c) shows the sizes $S$ of the giant component as functions of $p$ for  AA, DL, and UA. We can find
that a large value of $\alpha$ always leads to a larger size $S$ of the giant component for three layers, which means that the robustness of the system can be improved by greatly restricting the interdependence across network layers with increasing $\alpha$.

\section{Conclusion}
The interdependence of real multilayer networks is generally weak in layer-to-layer interactions, where the failure of one node usually does not result in failures of interdependent nodes across all network layers. In this paper, we have examined the percolation process and the robustness of a multilayer network when the interdependence of nodes across networks is weak. We reveal that the avalanche process of the whole system can be essentially decomposed into two microscopic cascading dynamics in terms of the propagation direction of failures: depth penetration and scope extension. Specifically, the former describes the propagation of failures across network layers and thus is regarded as ``cross-layer cascading'', while the latter describes the propagation of failures inside a network layer and thus is regarded as ``inner-layer cascading''. With the co-action of the two cascading dynamics, a multilayer network can disintegrate via first- or second-order percolation transitions in the case of initial failures, where the interdependence across network layers plays important roles in determining the percolation behaviors of the system. When the interdependence of network layers is weak, the failures of nodes can neither penetrate into deep network layers nor cause great destructiveness to their interdependent replicas, which inhibits the spread of failure and makes the system percolate via a second-order percolation transition. When the interdependence of network layers is strong, the failures of nodes can penetrate into deep network layers in a cascading manner and spread with a broad scope through various network layers, which thus makes the system collapse abruptly. These results prove that the process of ``cross-layer cascading'' dominates over the process of ``inner-layer cascading'' and plays a crucial role in determining the robustness of a multilayer system.

The present work essentially reveals the complexity of cascading failures, and previous works ignoring the weak interdependence may underestimate the complexity. Specifically, the cascading dynamics that occur across network layers cannot be produced in a strong layer-to-layer interdependence of multilayer networks. Our work not only offers a new understanding of the cascading failure dynamics of multilayer networks but also implies that the strength of interdependence can be exploited to enhance the robustness of multilayer networks against cascading failures. This can be especially meaningful in the engineering design of complex infrastructure systems that are intrinsically multilayer structured. Furthermore, our method also provides insights for the intervention of cascading failures in a multilayer network and evidences the idea that imposing restrictions on  ``cross-layer cascading'' can restrain the spread of failures more effectively.

\ack
This work was supported by the National Natural Science Foundation of China under Grant No.~61773148.

\section*{References}
\bibliography{reference}

\providecommand{\newblock}{}
\begin{thebibliography}{10}
\expandafter\ifx\csname url\endcsname\relax
  \def\url#1{{\tt #1}}\fi
\expandafter\ifx\csname urlprefix\endcsname\relax\def\urlprefix{URL }\fi
\providecommand{\eprint}[2][]{\url{#2}}

\bibitem{Klosik:2017}
Klosik D~F, Grimbs A, Bornholdt S and Hütt M~T 2017 {\em Nat. Commun.\/} {\bf
  8} 534

\bibitem{Ouyang:2014}
Ouyang M 2014 {\em Reliability engineering \& System safety\/} {\bf 121} 43--60

\bibitem{Rinaldi:2001}
Rinaldi S~M, Peerenboom J~P and Kelly T~K 2001 {\em IEEE Control Systems\/}
  {\bf 21} 11--25

\bibitem{Radicchi:2015}
Radicchi F 2015 {\em Nat. Phys.\/} {\bf 11} 597--602

\bibitem{BPPSH:2010}
Buldyrev S~V, Parshani R, Paul G, Stanley H~E and Havlin S 2010 {\em Nature
  (London)\/} {\bf 464} 1025--1028

\bibitem{KABGP:2014}
Kivel\"{a} M, Arenas A, Barthelemy M, Gleeson J~P, Moreno Y and Porter M~A 2014
  {\em J. Complex Net.\/} {\bf 2} 203--271

\bibitem{Gao:2011}
Gao J, Buldyrev S~V, Havlin S and Stanley H~E 2011 {\em Phys. Rev. Lett.\/}
  {\bf 107} 195701

\bibitem{Gao:2012}
Gao J, Buldyrev S~V, Stanley H~E and Havlin S 2012 {\em Nat. Phys.\/} {\bf 8}
  40--48

\bibitem{Baxter248701:2012}
Baxter G~J, Dorogovtsev S~N, Goltsev A~V and Mendes J~F~F 2012 {\em Phys. Rev.
  Lett.\/} {\bf 109}(24) 248701

\bibitem{Kesten:1982}
Kesten H 1982 {\em {Percolation Theory for Mathematicians}\/} (Boston:
  Birkh\"{a}user Press)

\bibitem{Stauffer:1992}
Stauffer D and Aharony A 1992 {\em Introduction to Percolation Theory\/}
  (London: Tailor \& Francis Press)

\bibitem{Son:2012}
Son S~W, Bizhani G, Christensen C, Grassberger P and Paczuski M 2012 {\em
  EPL\/} {\bf 97} 16006

\bibitem{Cellai:2013}
Cellai D, L\'opez E, Zhou J, Gleeson J~P and Bianconi G 2013 {\em Phys. Rev.
  E\/} {\bf 88} 052811

\bibitem{Hu:2013}
Hu Y, Zhou D, Zhang R, Han Z, Rozenblat C and Havlin S 2013 {\em Phys. Rev.
  E\/} {\bf 88} 052805

\bibitem{Kleineberg:2016}
Kleineberg K~K, Boguñá M, Ángeles Serrano M and Papadopoulos F 2016 {\em
  Nature Physics\/} {\bf 12} 1076–1081

\bibitem{Kleineberg:2017}
Kleineberg K~K, Buzna L, Papadopoulos F, Bogu\~n\'a M and Serrano M~A 2017 {\em
  Phys. Rev. Lett.\/} {\bf 118}(21) 218301

\bibitem{Faqeeh:2018}
Faqeeh A, Osat S and Radicchi F 2018 {\em Phys. Rev. Lett.\/} {\bf 121}(9)
  098301

\bibitem{Min:2014}
Min B, Yi S~D, Lee K~M and Goh K~I 2014 {\em Phys. Rev. E\/} {\bf 89} 042811

\bibitem{Parshani:2011}
Parshani R, Rozenblat C, Ietri D, C D and S H 2011 {\em EPL\/} {\bf 92} 68002

\bibitem{Valdez:2013}
Valdez L~D, Macri P~A, Stanley H~E and Braunstein L~A 2013 {\em Phys. Rev. E\/}
  {\bf 88} 050803

\bibitem{PBH:2010}
Parshani R, Buldyrev S~V and Havlin S 2010 {\em Phys. Rev. Lett.\/} {\bf 105}
  048701

\bibitem{Schneider:2013}
Schneider C~M, Yazdani N, Araújo N~A~M, Havlin S and Herrmann H~J 2013 {\em
  Scientific Reports\/} {\bf 3} 1969

\bibitem{Valdez:2014}
Valdez L~D, Macri P~A and Braunstein L~A 2014 {\em Journal of Physics A:
  Mathematical and Theoretical\/} {\bf 47} 055002

\bibitem{Berezin:2015}
Berezin Y, Bashan A, Danziger M~M, Li D and Havlin S 2015 {\em Sci. Rep.\/}
  {\bf 5} 8934

\bibitem{Bashan:2013}
Bashan A, Berezin Y, Buldyrev S~V and Havlin S 2013 {\em Nat. Phys.\/} {\bf 9}
  667--672

\bibitem{Danziger:2014}
Danziger M~M, Bashan A, Berezin Y and Havlin S 2014 {\em J. Complex Net.\/}
  {\bf 2} 460--474

\bibitem{Shekhtman:2014}
Shekhtman L~M, Berezin Y, Danziger M~M and Havlin S 2014 {\em Phys. Rev. E\/}
  {\bf 90} 012809

\bibitem{Shao:2014}
Shao S, Huang X, Stanley H~E and Havlin S 2014 {\em Phys. Rev. E\/} {\bf 89}
  032812

\bibitem{Huang:2013}
Huang X, Shao S, Wang H, Buldyrev S~V, Stanley H~E and Havlin S 2013 {\em
  EPL\/} {\bf 101} 18002

\bibitem{Emmerich:2014}
Emmerich T, Bunde A and Havlin S 2014 {\em Phys. Rev. E\/} {\bf 89} 062806

\bibitem{Yuan:2015}
Yuan X, Shao S, Stanley H~E and Havlin S 2015 {\em Phys. Rev. E\/} {\bf 92}
  032122

\bibitem{Hu:2017PNAS}
Yuan X, Hu Y, Stanley H~E and Havlin S 2017 {\em Proc. Nat. Acad. Sci. (USA)\/}
  {\bf 114} 3311--3315

\bibitem{Shaw:2010}
Shaw L~B and Schwartz I~B 2010 {\em Phys. Rev. E\/} {\bf 81}(4) 046120

\bibitem{Zhao:2014}
Zhao D, Wang L, Li S, Wang Z, Wang L and Gao B 2014 {\em PloS One\/} (9)
  e112018

\bibitem{NA:2015}
Nacher J~C and Akutsu T 2015 {\em Phys. Rev. E\/} {\bf 91}(1) 012826

\bibitem{WangZexun}
Wang Z, Zhou D and Hu Y 2018 {\em Phys. Rev. E\/} {\bf 97}(3) 032306

\bibitem{Liu:2019}
Liu R~R, Jia C~X and Lai Y~C 2019 {\em New Journal of Physics\/} {\bf 21}
  045002

\bibitem{Bashan2011}
Bashan A and Havlin S 2011 {\em Journal of Statistical Physics\/} {\bf 145}
  686--695 ISSN 1572-9613

\bibitem{Bashan2011b}
Bashan A, Parshani R and Havlin S 2011 {\em Phys. Rev. E\/} {\bf 83}(5) 051127

\bibitem{Liming:2013}
Li M, Liu R~R, Jia C~X and Wang B~H 2013 {\em New Journal of Physics\/} {\bf
  15} 093013 \urlprefix\url{http://stacks.iop.org/1367-2630/15/i=9/a=093013}

\bibitem{Azimi:2014}
Azimi-Tafreshi N, G\'omez-Garde\~nes J and Dorogovtsev S~N 2014 {\em Phys. Rev.
  E\/} {\bf 90} 032816

\bibitem{Baxter:2014}
Baxter G~J, Dorogovtsev S~N, Mendes J~F~F and Cellai D 2014 {\em Phys. Rev.
  E\/} {\bf 89} 042801

\bibitem{Radicchi:2017}
Radicchi F and Bianconi G 2017 {\em Phys. Rev. X\/} {\bf 7} 011013

\bibitem{LESL:2018}
Liu R~R, Eisenberg D~A, Seager T~P and Lai Y~C 2018 {\em Sci. Rep.\/} {\bf 8}
  2111

\bibitem{Liu:2016B}
Liu R~R, Li M and Jia C~X 2016 {\em Sci. Rep.\/} {\bf 6} 35352

\bibitem{Molloy:1995}
Molloy M and Reed B 1995 {\em Algorithms\/} {\bf 6} 161--180

\bibitem{Son2011}
Son S~W, Bizhani G, Christensen C, Grassberger P and Paczuski M 2011 {\em
  Epl\/} {\bf 97} 16006

\bibitem{Feng:2015}
Feng L, Monterola C~P and Hu Y 2015 {\em New Journal of Physics\/} {\bf 17}
  063025

\bibitem{CEAH:2000}
Cohen R, Erez K, ben Avraham D and Havlin S 2000 {\em Phys. Rev. Lett.\/} {\bf
  85} 4626--4629

\bibitem{CNSW:2000}
Callaway D~S, Newman M~E~J, Strogatz S~H and Watts D~J 2000 {\em Phys. Rev.
  Lett.\/} {\bf 85} 5468--5471

\end{thebibliography}
\end{document}